\documentclass[a4paper,fleqn,usenatbib]{mnras}
\usepackage{newtxtext,newtxmath}
\usepackage[T1]{fontenc}
\usepackage{ae,aecompl}

\usepackage{graphicx}	
\usepackage{amsmath}	
\usepackage{amssymb}	
\usepackage{ulem}
\usepackage{hyperref}

\newcommand{\Mpch}{\,Mpc/$h$}	

\title[Partitioning basins]{Partitioning the universe into gravitational basins using the cosmic velocity field}

\author[ Dupuy et al.]{Alexandra Dupuy$^{1}$, Helene M. Courtois$^{1}$, Florent Dupont$^{2}$,  Florence Denis$^{2}$, 
\newauthor Romain Graziani$^{1}$, Yannick Copin$^{1}$, Daniel Pomar\`ede$^{3}$, Noam Libeskind$^{1,4}$, 
\newauthor Edoardo Carlesi$^{4}$, Brent Tully$^{5}$, Daniel Guinet$^{1}$\\
$^{1}$University of Lyon, UCB Lyon 1, CNRS/IN2P3, IPN Lyon, 69622 Villeurbanne, France\\
$^{2}$Univ Lyon, LIRIS, UMR 5205 CNRS, Universit\'e Claude Bernard Lyon 1, 43 bd du 11 Novembre 1918, 69622 Villeurbanne CEDEX, France\\
$^{3}$Institut de Recherche sur les Lois Fondamentales de l'Univers, CEA, Universit\'e Paris-Saclay, 91191 Gif-sur-Yvette, France\\
$^{4}$Leibniz--Institut f\"{u}r Astrophysik, Potsdam, An der Sternwarte 16, D--14482 Potsdam, Germany\\
$^{5}$Institute for Astronomy, University of Hawaii, 2680 Woodlawn Drive, Honolulu, HI 96822, USA\\
}

\date{Accepted XXX. Received YYY; in original form ZZZ}

\pubyear{2018}

\begin{document}
\label{firstpage}
\pagerange{\pageref{firstpage}--\pageref{lastpage}}
\maketitle

\begin{abstract}
This letter presents a new approach using the cosmic peculiar velocity field to characterize the morphology and size of large scale structures in the local Universe. The algorithm developed uses the three-dimensional peculiar velocity field to compute flow lines, or streamlines. The local Universe is then partitioned into volumes corresponding to gravitational basins, also called watersheds, among the different end-points of the velocity flow lines. This new methodology is first tested on numerical cosmological simulations, used as benchmark for the method, and then applied to the {\it Cosmic-Flows} project observational data in order to to pay particular attention to the nearby superclusters including ours. More extensive tests on both simulated and observational data will be discussed in an accompanying paper.
\end{abstract}




\section{Introduction}

The current cosmological paradigm, known as the $\Lambda$CDM model, lends itself to a hierarchical classification scheme, wherein small dwarf galaxies merge to form galaxies like the Milky Way, which are bound together into galaxy groups, which are in turn part of galaxy clusters. Such structures exist within the {\it cosmic web} \citep{Bond:1996aa}, a multi-scale network of clusters, filaments, sheets and voids which constitutes the large scale distribution of matter in the universe. The term supercluster is used to define an ensemble of such knots, filaments, walls,  and voids.

Hence, the measured large-scale distribution of matter in the universe can be used to measure constrain our cosmological model. Upcoming observational (such as Euclid, DESI or LSST) and computational (such as MultiDark) projects are based upon this idea that the study of the distribution (and motion) of galaxies can be a probe of cosmology. However, to infer this and cosmological parameters directly from the distribution of galaxies, a robust definition and quantification of the large-scale structures are required. 

Many methods for classifying the local large-scale structures using galaxy redshift surveys have been developped in recent years. For example \cite{Sousbie:2008aa, Sousbie:2011aa} introduced the {\it DisPerSE} algorithm which identifies the skeleton of clusters, filaments, walls and voids directly from the distribution of galaxies in observational catalogs such as the Sloan Digitized Sky survey (SDSS). {\it SpineWeb} \citep{Aragon-Calvo:2007aa, Aragon-Calvo:2010aa} is a method based on the watershed segmentation of the density field derived from the observed galaxy redshift distribution, allowing for a multiscale morphology filter to identify clusters, filaments and walls. Similarly, {\it Bisous} \citep{Tempel:2014aa} studies the redshift galaxy distribution to find patterns and regularities in galaxy filaments. A new exciting approach has recently been suggested by \cite{Leclercq:2017aa}, who advocates using the gravitational velocity field derived from redshift surveys such as SDSS to identify structures.

On the other hand, anticipating the exponential growth of kinematical (peculiar velocity) galaxy data, like $Cosmic-Flows$, 6DF, TAIPAN, WALLABY, a number of  methods were developed based on computing the tidal tensor field (Hessian of the gravitational potential)  to classify large scale structures (LSS) \citep{Hahn:2007aa,Forero-Romero:2009aa}.  Similarly \cite{Hoffman:2012aa} used the  velocity shear tensor to classify volumes as knots, filaments, sheets, voids, according to the tensor's eigenvalues.

A third category is combining redshift and peculiar velocity data. For example, $NEXUS$  \citep{Cautun:2013aa} is a method which uses the density field, the velocity divergence and the velocity shear fields to identify automatically volumes of space as cosmic web structures. See \cite{Libeskind:2018aa} for a comprehensive review and comparison between all these, and other, methods. 

Recently, \cite{Tully:2014aa} suggested that one could additionally employ kinematics to define superclusters, as non-virialized objects that will dissipate with the expansion. Superclusters are not gravitationally bound -- their sheer size and existence within an expanding Universe prohibits that, and so these previous methods do not fully capture the scale of these structures. However if the expansion of the Universe is instantaneously frozen, regions of space show gravitationally induced coherent inward motions, known as ``Basins of attraction'' (BoA). This simple definition allows also one to identify regions that are dual to basins of attraction -  basins of repulsion (BoR), namely volumes of space with gravitationally coherent outward motions such as the dipole repeller \citep{Hoffman:2017aa} and the CMB Cold Spot repeller \citep{Courtois:2017aa}. In order to identify such structures we must observe the gravitational velocity (also called peculiar velocity) of a representative sample of galaxies in a large enough region of space.

This article proposes a different approach to all current methods of cosmic web classifications by focusing on this kinematical new definition of superclusters. This new tool can prove superclusters as robust probes of the cosmology when applied to larger upcoming redshift surveys spanning a good fraction of the universe age, like Euclid.
The article is organized as follows: in Section \ref{sec:method} the methodology is explained, in Section \ref{sec:tests} it is it is tested on $\Lambda$CDM cosmological simulations and in Section \ref{sec:observations} it is used on observed peculiar velocity data.

\section{Methodology}
\label{sec:method}
\subsection{Streamlines}

A stream- or flow line, in a time-independent velocity field, represents the curve that is always tangent to the local value of the velocity field. It can be parametrized in terms of a coordinate $\tau$, which represents the position along the streamline $s$, as follows:

\begin{equation}
\label{eq:streamline1}
\frac{d\vec{s}}{d\tau} = \vec{v}(\vec{s}),
\end{equation}
where $\vec{s}(\tau=0) = \vec{s}_0$ corresponds to the initial condition and is the {\it seed point} of the streamline. The streamline $\vec{s}(\tau)$ is obtained by integrating equation \ref{eq:streamline1} over $\tau$. Thus, a set of streamline points $\{\vec{s}_i\}_{i \in [0,n]}$ is generated, where $n$ is the number of integration steps $\Delta\tau$. 

A streamline is therefore obtained step by step by starting from a seed point (each voxel) and integrating spatially the components of the velocity field. For clarity purposes, the end of a streamline, i.e the location $\vec{s}_n$ at which the computation of $\vec{s}$ is stopped, is noted $\vec{s}_\mathrm{stop}$ and called the ``stop point'' throughout this article.

Streamlines converge towards the critical points of the velocity field, which correspond to the positions at which the velocity is zero, i.e, the stop point. In a peculiar velocity field, we define such locations as attractors (centers of attracting regions) or repellers (centers of emptying regions). In order to visualize these structures, one can specify the direction in which streamlines are generated. On the one hand, a streamline generated in the {\it forward} direction is obtained by integrating the velocity field $\vec{v}$ from a given seed point, and the stop point corresponds to a sink, or attractor. On the other hand, a streamline generated in the {\it backward} direction is achieved by integrating $-1 \times \vec{v}$ from a given seed point, allowing us to visualize emptying volumes.This is equivalent to setting the step $\Delta\tau$ negative.

In the study presented here, the velocity field is discretised in a regular cubic grid. Each voxel (cell) is used to generate a streamline. Furthermore, streamlines are generated by the use of the \texttt{Streamtracer} function of the VTK library \citep{VTK:2006}. The integration step $\Delta\tau$ is set to the resolution of the computation grid (size of a single voxel), as this value is the typical step size which allows to obtain converged results and well-defined basins' borders (see tests in accompanying paper). 
Streamlines are generated by the use of the fourth order Runge-Kutta method (RK4). 
The computation of a streamline is stopped if the maximal number of integrations, set beforehand, is reached. This parameters controls the length of streamlines. Tests described in the accompanying paper show that results converge with a typical length of streamlines equivalent approximately to the size of the box in which the velocity field is computed.
The velocity field is interpolated by the VTK \texttt{Streamtracer} function in order to be evaluated at the positions of the streamlines' points at each integration step. This interpolation and parameters related to the computation of streamlines will be studied in the accompanying paper along other studies like grid resolution effects.

\subsection{Identification of basins}

Once streamlines are computed as described in the previous section, we apply the following algorithm in order to partition regions of space into basins of attraction and repulsion.

Consider a peculiar velocity field $\vec{v}$ discretised into a regular, cubic three-dimensional grid of size $N^3$. $N^3$ thus represents the dimensionality of the discretised velocity field and sets the scale on which the velocity field is computed. A three-dimensional grid $M_\mathrm{stop}$ (also of dimensionality $N^3$) is constructed by counting the number of streamlines whose final destination $\vec{s}_\mathrm{stop}$ is in a given cell.  Local maxima of $M_\mathrm{stop}$ correspond thus to the locations of the center's of attractors and repellers (depending on the direction of integration). A local maximum of $M_\mathrm{stop}$ is detected if the number of streamlines which end in a given cell is (a) greater than the neighbouring cells and (b) greater than a threshold value. 
This procedure results in $N_{\rm max}$ local maxima $\{\vec{x}_{1}, \vec{x}_{2}, ... , \vec{x}_{N_{\rm max}}\}$ labeled (arbitrarily) as $\{l_{1}, l_{2}, ... , l_{N_{\rm max}}\} \neq 0$. Therefore, voxels in $M_\mathrm{stop}$ are either given a value of 0, if they are not local maxima, or $\{l_{i}\}$ if they are.

Let us now consider another three-dimensional grid $B$ (also of dimensionality $N^3$). The purpose of $B$ is to assign each voxel to a given basin center, $\{\vec{x}_{i}\}$. Each voxel that is not a local maxima  is allocated to the basin in which its streamline stops. Note that in our algorithm each voxel represents the starting point of a streamline which ends at one of the local maxima $\{\vec{x}_{i}\}$. In other words voxels are all (1) starting sites for streamlines, but may also (2) host other streamlines or (3) be the end points of a streamline(s). It is the linking of the starting points voxels with the stopping point voxels (local maxima) which define the basins. 

\begin{figure}
\includegraphics[width=.4\textwidth,angle=-00]{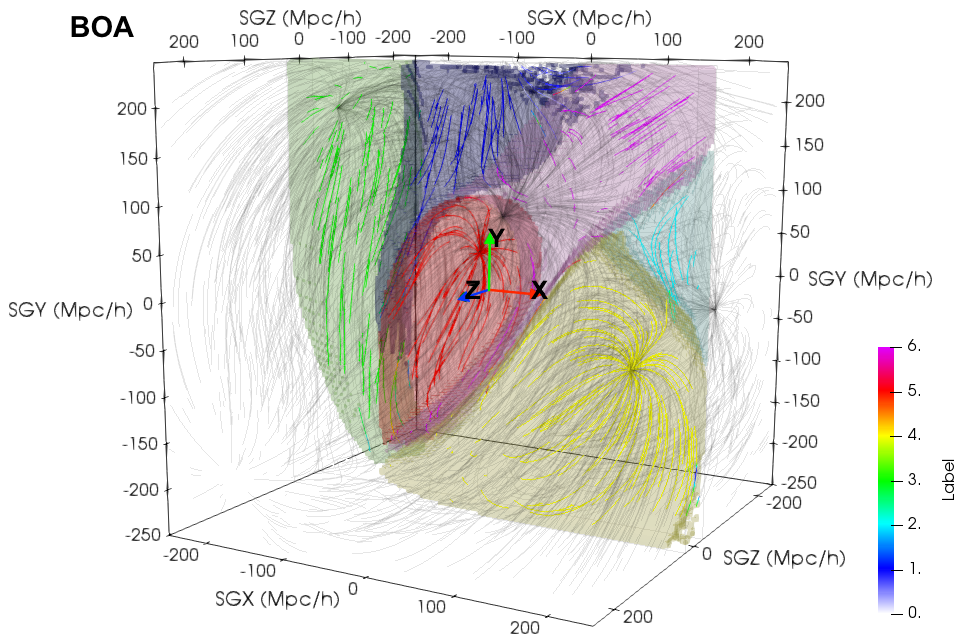}\\
\caption{Automatic segmentation of a volume of space where six attractors are localized. Only the central basin (in red) is totally segmented since it is
surrounded by other attractors. Attractors close to the volume boundaries cannot be fully segmented (no backdrop). Color code corresponds to the "voxels labelling"as explained in the text.}
\label{test_BOA}
\end{figure}
 Figure \ref{test_BOA} displays the result of the segmentation algorithm on a simple distribution of 6 artificial attractors in a volume.

\section{Tests on constrained simulations}
\label{sec:tests}

The method described above is first tested in an ideal setting, namely by using numerical cosmological simulations. The simulation considered in this section is an $N$-body $\Lambda$CDM simulation with $256^3$ dark matter particles of mass $6.57 \times 10^{11}$ M$_\odot/h$, which has been run with the simulation code GADGET-2 \citep{Springel:2005aa} as part of the Constrained Local UniversE Simulations (CLUES) and Cosmicflows collaborations \citep{Yepes:2009aa,Gottloeber:2010aa,Courtois:2012aa,Sorce:2016aa}. This simulation provides the density field $\rho_\mathrm{simu}$ and the three cartesian components of the peculiar velocity field $\vec{v}_\mathrm{simu}$. These two fields have been computed in a box of dimensions $256^3$ and of width 500\Mpch. A $\Lambda$CDM  Planck 2013 \citep{Ade:2013zuv} cosmology has been assumed and the {\it Cosmicflows-2} catalog \citep[CF2,][]{Tully:2013aa} has been used to generate the initial conditions using the constrained realization technique \citep{Hoffman:1992aa} and the reverse Zeldovich approximation \citep{Doumler:2013aa}. 

We can compute an image of the streamlines, i.e.an histogram of the number density of streamlines that cross a voxel in space.

We basically count the number of streamlines that intersect the voxels.
The matrix $S$ thus includes values ranging from 1 (only one streamline passes through a voxel, i.e the streamline starting from this voxel) to a huge number (voxels located at attractors or repellers positions, i.e locations to which lots of streamlines converge).

\begin{figure*}
\includegraphics[width=0.6\textwidth,angle=-00]{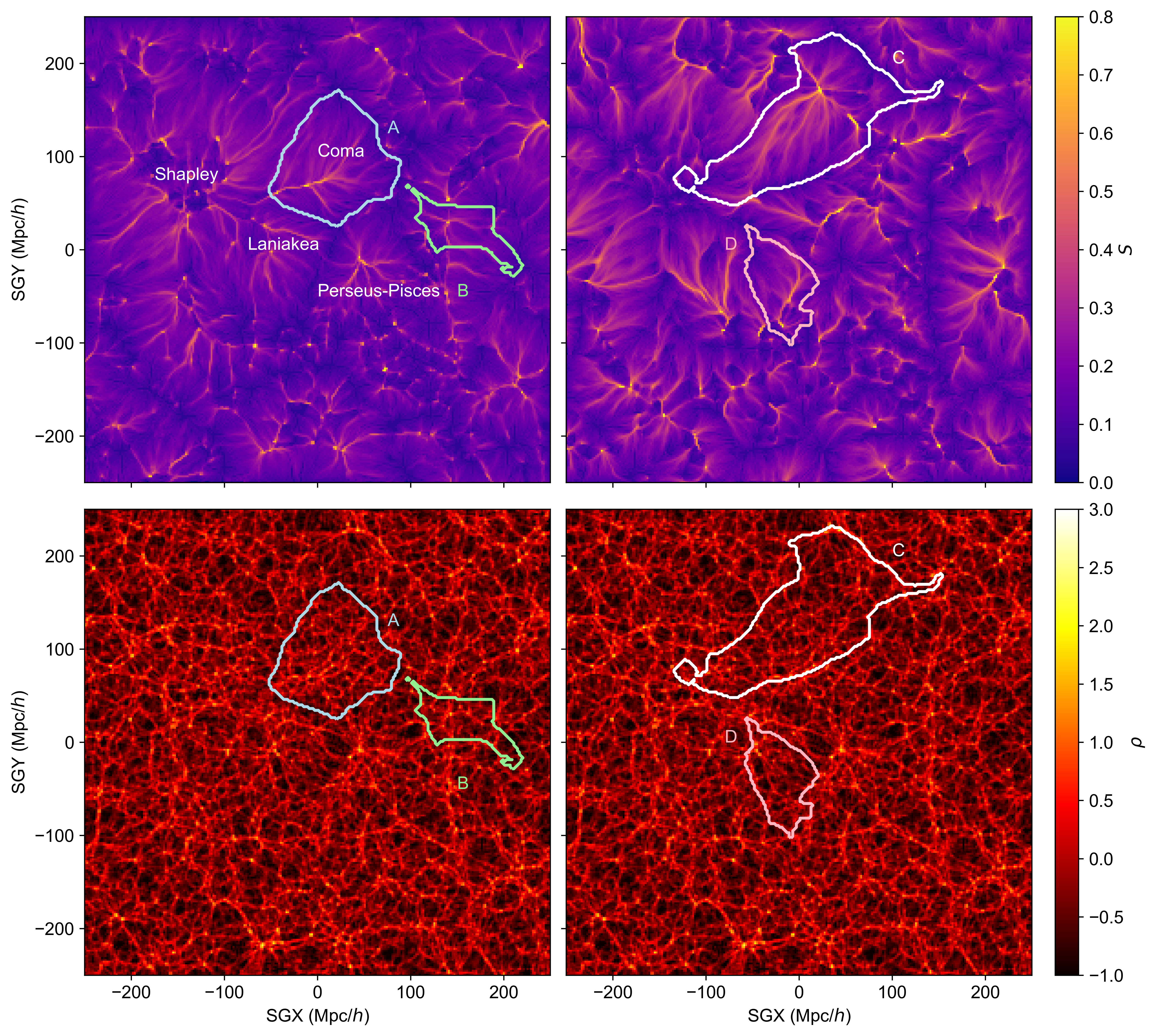}
\caption{Top panels show the "image", i.e, density, of streamlines in supergalactic plane XY, in a $\Lambda$CDM simulation whose initial conditions were constrained by the Cosmic-flows-2 data. The top left panel shows streamlines computed forward on the $v$ velocity field, while top right panel shows streamlines computed backwards, i.e.on $-1 \times v$ velocity field. Bottom panels show the corresponding matter density field $\rho$. Four examples of automatically segmented basins of attractions A and B (left column) and repulsion C and D (right column) are shown respectively with a blue, green, white and pink boundary solid lines.}
\label{simu_BOA}
\end{figure*}

The figure \ref{simu_BOA} presents four panels in supergalactic coordinates SGX-SGY at SGZ = 0 Mpc/$h$. The "image", i.e.density, of streamlines is displayed in the two top panels. On the left for the forward velocity field, and on the right for the backward velocity field. Two examples of basins of attraction (A and B) and repulsion (C and D), were computed with the segmentation methodology. The bottom panels display those superclusters and emptying regions boundary outlines as solid perimeters on top of the matter density $\rho$ of the simulation. The usual filamentary structure of the matter density can be observed in red, with the highest peaks, i.e.clusters and filaments, in yellow. It is quite striking that from the density alone the boundaries of A and B superclusters, and C and D emptying regions, cannot be identified; however they appear very clearly as well-defined large scale structures in the images of streamlines.

\begin{figure}
\includegraphics[width=0.45\textwidth,angle=-00]{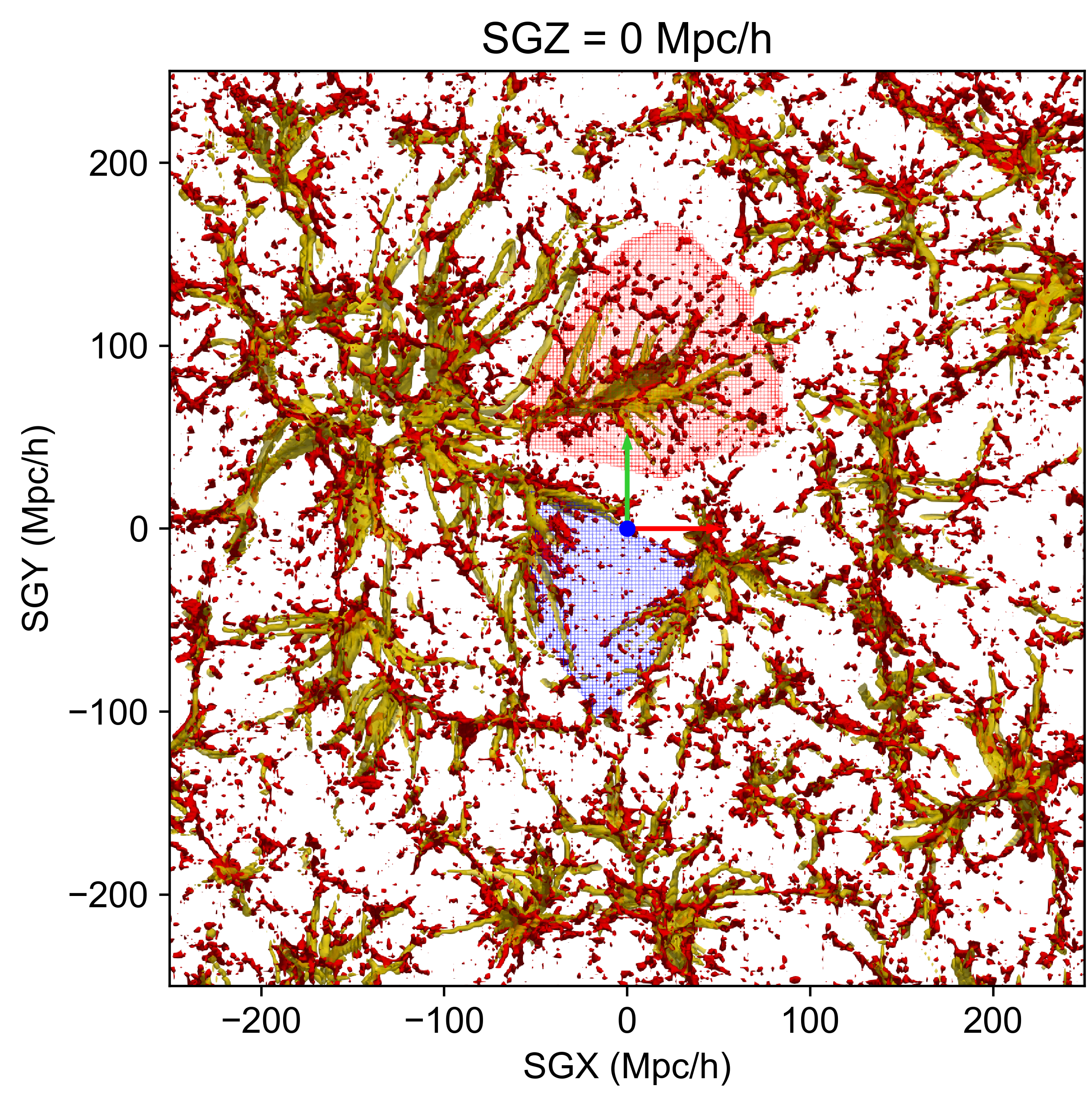}
\caption{This animated figure (see \url{https://vimeo.com/305959931/b85e36f10a}) displays the relation between high peaks in the matter density (in red) and the image of streamlines computed from the peculiar velocity field, in a $\Lambda$CDM simulation with initial conditions constrained on the Local Universe. The regions of high streamlines density connects regions of high matter density. Two automatically partitoned volumes are shown; in red mesh :  the one corresponding to a simulated Coma supercluster, in blue mesh a basin of repulsion.}
\label{rho}
\end{figure}

The algorithm finds 87 basins of attractions and 76 basins of repulsion in the $(500$ Mpc/$h)^3$ simulated local universe. This number is related to the voxel size; the grid resolution used here is the "linear" scale for the growth of structures. Peculiar velocities in Cosmic-flows catalogs do not probe non-linear scales typically below 4 Mpc/$h$. As some example, the simulation predicts local superclusters volumes as:  Laniakea $5\times10^5 ($Mpc/$h)^3 $, Coma supercluster $1\times10^6\ ($Mpc/$h)^3 $ (in red on Figure \ref{rho}), Perseus-Pisces supercluster $7\times10^5 ($Mpc/$h)^3$.

Once a supercluster size and location are kinematically defined, one can compute its total mass. Similarly it is possible to compute the mass of emptying regions. The size of a supercluster varies from $250-500 \times10^3 ($Mpc/$h)^3$, while emptying regions cover volumes of $250-700 \times10^3 ($Mpc/$h)^3$. The typical mass of a supercluster is $5\times10^{16} $M$\odot/h$. \\

On Figure \ref{rho} (see animation here : \url{https://vimeo.com/305959931/b85e36f10a}) can be appreciated the density field of the $\Lambda$CDM simulation as red isosurfaces, and the image (or density) of streamlines derived from the peculiar velocity field of the same simulation as yellow isosurfaces. Two gravitational basins are displayed as colored filled contours: one basin of attraction in red (which corresponds to the Coma superclusters, see explanations above), and one basin of repulsion in blue. One can easily see that the isosurfaces of the image of streamlines connect the densities $\rho$ of the simulation belonging to the same gravitational basin.

\section{Testing nearby superclusters}
\label{sec:observations}

\begin{figure*}
\includegraphics[width=0.76\textwidth,angle=-00]{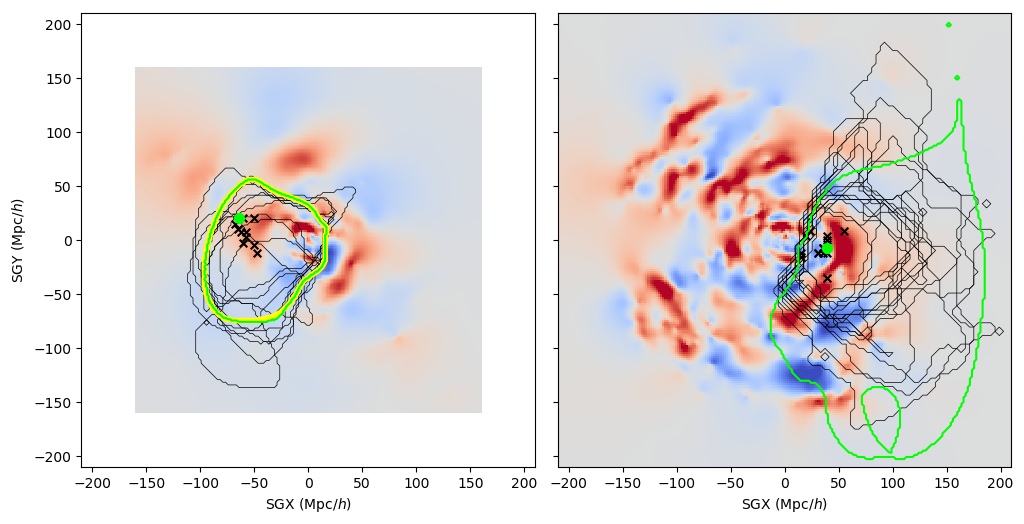}
\caption{Automatic segmentation of the Laniakea (left) and Perseus-Pisces (right) superclusters using CF2 and CF3 data respectively, displayed in the SGX-SGY plane centered on $\mathrm{SGZ} = 0$ Mpc/$h$. The colormaps represents the corresponding reconstructed CF2 and CF3 overdensity fields, with underdensities in highlighted in blue, and regions of high density in red. Locations of attractors and frontiers of the superclusters obtained from the local CF2 and full CF3 velocity fields are represented by green circles and solid lines. Boundaries derived by the use of 20 constrained realizations of each Cosmicflows dataset are displayed as black solid lines, while corresponding attractors are located as black crosses. The yellow solid line corresponds to the borders of our home supercluster Laniakea as defined in Tully et al. (2014).}
\label{BOAcosmicflows}
\end{figure*}

In this section, the segmentation methodology is applied to velocity fields reconstructed from observational data, i.e the {\it Cosmicflows-2} \citep[CF2,][]{Tully:2013aa} and {\it Cosmicflows-3} \citep[CF3,][]{Tully:2016aa} catalogs. The velocity fields computed from the CF2 and CF3 datasets allow to study two nearby superclusters: Laniakea and Perseus-Pisces respectively.


The CF2 reconstructed velocity field is computed using the Wiener Filter algorithm \citep{Zaroubi:1999aa}, in a box of dimensions $256^3$ and of width 320 Mpc/$h$. The left panel of Figure \ref{BOAcosmicflows} displays as a green solid line the boundary of Laniakea, resulting from the segmentation of the {\it local} CF2 three-dimensional velocity field, after filtering out the tidal components caused by distant structures. The methodology applied here to separate the local and tidal flows is the same as the one used in \cite{Tully:2014aa} in order to define our home supercluster. Indeed, adjacent large scale structures such as Coma, Perseus-Pisces, Shapley and Hercules cannot be fully captured due to the lack of data coverage in the CF2 catalog. The local flow is then extracted by computing the velocity field from the overdensity field contained in a sphere of 100 Mpc/$h$ radius centered on the Norma supercluster and ignoring all the structures outside.  For the sake of comparison, the manually found boundary of the Laniakea supercluster as defined in \cite{Tully:2014aa} is shown in the left panel of Figure \ref{BOAcosmicflows} as a yellow solid line. These two frontiers of the supercluster look similar with respectively enclosed volumes of $2.3 \times 10^6$ (Mpc/$h$)$^3$ (automatic -- green solid line) and $1.7 \times 10^6$ (Mpc/$h$)$^3$ (manual -- yellow). The data, definition and methodology used to identify Laniakea are the same as in \cite{Tully:2014aa}, however the algorithm finds now automatically gravitational basins, instead of manually.


The velocity field reconstructed from the CF3 data is derived in a box of dimensions $256^3$ and of width 500 Mpc/$h$, by forward-modeling the data and computing a set of about a thousand constrained realizations converging with a Monte-Carlo-Markov chain, as proposed by \cite{Graziani:2018aa}, instead of deriving just one single Wiener Filter reconstruction as done for CF2. In this case, the segmentation methodology is applied to the {\it full} velocity field, i.e the local and tidal components are not separated. On the right panel of Figure \ref{BOAcosmicflows} can be appreciated the frontier of our neighbouring supercluster Perseus-Pisces displayed as a green solid line. This automatically segmented region encloses a volume of $8.8 \times 10^6($Mpc/$h$)$^3$. This volume is however overestimated, as the basin of Perseus-Pisces is not well-defined in the outer region due to the lack of data in this direction. In order to better define the limits of this gravitational basin, one would need to observe more distant structures behind Perseus-Pisces. This letter will be soon followed by a more detailed analysis of the automated segmentation derived from the CF3 reconstructed velocity field. The partitioning analysis will provide identifications of other nearby superclusters like Hercules, Shapley, etc., as well as errors estimates. 

It is to be noted that although the Cosmicflows datasets, especially CF2, are very coarse, these volumes, as seen with tests in the previous section, are realistic for a $\Lambda$CDM universe. For example, Laniakea's and Perseus-Pisces' volumes are found to be, in a $\Lambda$CDM constrained simulation,  $5 \times 10^5$ (Mpc/$h$)$^3$ and $7 \times 10^5$ (Mpc/$h$)$^3$ respectively (see Section \ref{sec:tests}).


This automated segmentation of volumes also brings the new capacity of estimating errors on supercluster sizes by the use of constrained realizations (CRs) of the reconstructed velocity fields. In addition to the CF2 and CF3 velocity fields, twenty realizations of each velocity field are considered throughout this section. Here, the segmentation algorithm is applied to the full velocity fields of the CRs, for both CF2 and CF3 datasets. The black crosses and black solid lines displayed in Figure \ref{BOAcosmicflows} represent the location of the central attractor and boundaries of the Laniakea (right) and Perseus-Pisces (left) superclusters obtained from the automated segmentation applied on 20 constrained realizations of the CF2 and CF3 velocity fields respectively. The Wiener Filter is by definition the mean over Constrained Realizations (CRs) of the local universe. We define the mean positions, and their associated error, of the basins as the mean positions over CRs (resp. the standard deviation). We obtain: Laniakea (-58$\pm$6, +6$\pm$10, +21$\pm$15) Mpc/$h$, Perseus-Pisces supercluster (33$\pm$10, -7$\pm$10, 31$\pm$9) Mpc/$h$.

In the case of CF2, the tidal and local flows have to be separated to identify Laniakea on observational data (see above). This methodology can lead to different velocity fields and thus different gravitational basins depending on the parameters considered. However, here,  this methodology has not been applied to the CF2 CRs. Applying the same approach to both data and CRs may lead to a better agreement between the two.

\section{Conclusions}

This letter proposes an automated way of partitioning the universe into gravitational basins, by constructing and following streamlines directly from the gravitational (peculiar) velocity field at a resolution above non-linearities (galaxy collisions, clusters of galaxies) and at a given instant in cosmic time.

As the Figures \ref{simu_BOA} and \ref{rho} explicitly show, it is difficult to identify basins of attraction or repulsion from the density field only. Moreover the full density field is not available from observations. This methodology based on the peculiar velocity field brings a new capacity to the field of identifying physically connected large scale structures.

The typical size of basins of attraction has been found to be a volume of the order $10^5-10^6 ($Mpc/$h)^3$ both in the observed and simulated $\Lambda$CDM universe. Figure \ref{BOAcosmicflows} shows the current results for the segmentation of the observed Local Universe. The Laniakea and Perseus-Pisces superclusters are easily identified by their basins of attractions, however, at the most distant scales, one can see that both CF2 and CF3 reconstructions are still clearly showing artefacts.

The present-day, current, velocity field is considered throughout this letter. Such a velocity field does not necessarily point along the direction that the actual mass transport takes place. Thus, even if all the flow lines converge to one point, it does not mean that all the mass associated to those flow lines will actually (when taking into account the time evolution of the velocity field) end up in that point. An analysis at other redshifts is included in the accompanying paper, allowing us to study the evolution of gravitational basins over time (basins' number and total mass).

As stated above, this automated segmentation method will be tested further in a more extensive article which will follow. A future goal would be to quantify the capacity of the segmentation method at deriving physical quantities usually unaccessible, such as the total mass enclosed in a gravitational basin, as seen in Section \ref{sec:tests}. 

\section*{Acknowledgements}
Authors acknowledge support from the Institut Universitaire de France and the CNES. AD and NIL acknowledges financial support of the Project IDEXLYON at the University of Lyon under the Investments for the Future Program (ANR-16-IDEX-0005). 

\bibliographystyle{mnras}
\bibliography{bibliothese}



\bsp	
\label{lastpage}
\end{document}